# The World Wide Web in the Face of Future Internet Architectures

Submitted in partial fulfillment of the requirements for

the degree of

Master of Science

in

Information Networking

Harshad J Shirwadkar

B.E., Computer Engineering, Pune Institute of Computer Technology
M.S., Information Networking, Carnegie Mellon University

Carnegie Mellon University
Pittsburgh, PA

May, 2016

# Acknowledgements

Working on master's thesis was my first real research experience. I thoroughly enjoyed it. It taught me to think about problems in a different way. I would like to thank everyone who directly or indirectly supported me throughout this wonderful journey.

I would like to express my deepest gratitude to my advisor, Professor Peter Steenkiste for his support and guidance throughout the duration of the project. He was always available to answer my questions and show me the right way. His continued support over the past two years has been crucial in this accomplishment. I would also like to acknowledge Professor Srinivasan Seshsan who served as my thesis reader and shared valuable feedback on the ideas.

I would like to thank the entire XIA research group for sharing valuable time to time feedback on my work. I loved being a part of the group. I would also like to thank Dan Barrett who helped me with understanding the huge codebase and with setting up various experiments. His door was always open whenever I ran into a trouble spot.

Also, I would like to thank Saurabh Kadekodi for many meaningful discussions that shaped ideas in the thesis.

Finally, I must express my very profound gratitude to my parents and my brother for providing me with their continuous encouragement and unfailing support. This work would not have been possible without them.

This research was funded in part by NSF under awards number CNS-1040801 and CNS-1345305.

# Abstract

The World Wide Web (WWW) has arguably been the most popular application of the Internet for years. Over a period of time, it has developed over the principles of host-centric IP internet. However, the limitations of today's host-centric IP internet have motivated many future internet architectures that are centered around alternate principals such as content and services.

In this thesis, we study the WWW and propose features needed by such clean slate future internet architectures that can benefit the WWW. The features that we propose are then implemented on eXpressive Internet Architecture (XIA) - a candidate future internet architecture.

Most of the clean slate architectures proposed so far revolve around an alternate principal giving rise to networking infrastructural styles such as content-centric-networking or service-oriented networking. XIA argues that elevating one principal above others limits the ability to communicate with the other principals. Thus, XIA inherently supports co-existence of multiple communication principals.

The WWW relies on a reliable transport layer for content delivery. We see how these coexisting principals cooperate to provide a new reliable content delivery architecture that offers content caching and reliable content transport as services. Despite being offered as services, we still maintain primary features offered by a content centric network such as in-network caching and content routing.

We define a new principal type that allows fetching content by human readable names rather than by cryptographically secure identifiers. Although human readable names are more convenient for the world wide web, they are more vulnerable to security threats than the cryptographically secure identifiers. We address authenticity, integrity issues raised by human readable names. We then define a security model that allows endpoints as well as in-network devices to perform integrity, authenticity checks in constant time.

To complete the story, we avoid the need of name lookup by defining URL scheme that directly address es the content. Based on these URL formats, the human readable identifiers and the new content delivery system proposed, we model the World Wide Web over XIA.

# Table of Contents

# List of Tables

# List of Figures

# 1
# World Wide Web

The World Wide Web is arguably the most important application of the Internet. The World Wide Web is an information space that allows exchanging information objects called web resources between hosts. It was invented by English scientist Tim Berners-Lee in 1989.

The web is so popular that the term is often used interchangeably with the Internet itself. However, these two are not the same. The Internet is a giant network of interconnected computers identified by IP address. Whereas, the web is a collection of web resources such as documents, videos, images that these interconnected computers can network. Essentially, the web runs *on top of* the Internet.

In this chapter we will take a closer look at the important components of the web.

## 1.1 Universal Resource Identifiers

The web resources are uniquely identified by something called as Universal Resource Identifiers or URIs. URI is a string of characters that can identify a web resource uniquely. This unique identification provides the network entities a way to identify and therefore *request* as well as *serve* a resource. RFC 2396 formally defined the format of URI. The definition was later refined by RFC 3986. The simplified definition of the URI is as follows.

```
URI              = scheme : hierarchical-part [ ? query ] [ # fragment ]
hierarchical-part   = // authority path
                    / path
```

- *Scheme:* Examples of popular schemes are HTTP, FTP, mailto etc.

- *Hierarchical Part:* Location of a web resource within some logical hierarchy. Often, this part is formed by combining the host (a registered name or an IPv4 address) and hierarchical path (similar to UNIX file system paths).

- *Query:* Traditionally consists of key-value pairs.

- *Fragment:* A character string that identifies a fragment in the resource. For example, a section in an article.

The example of a URI given in RFC 3986 is as follows.

```
foo://example.com:8042/over/there?name=ferret#nose
\_/   \______________/\________/ \________/ \__/
 |           |             |          |       |
scheme    authority       path      query  fragment
```

URL (Universal Resource Locator) is the most commonly used form of URI in the World Wide Web. In addition to uniquely identifying a web resource, URL also provides a way to locate the resource. Essentially, URL identifies a web resource by its network location.

## 1.2 Hypertext Transfer Protocol

The primary method used for publishing and retrieving these web resources is Hypertext Transfer Protocol (HTTP). Although HTTP is one of many Internet communication protocols, the web resources are usually accessed via HTTP. HTTP is a request-response type of protocol. The HTTP clients or web clients request resources by sending a `HTTP GET` request to entities serving these resources. The entities that serve the web resources are called HTTP servers or Web Servers.

### 1.2.1 HTTP Session

HTTP Session is a sequence of request-response transactions. The HTTP client initiates a reliable transport session (a TCP session) with a HTTP server listening on a particular predefined port. The port number used by HTTP servers is usually 80. Once the session has been established, the client then sends a HTTP request to the server. Server responds back with a status line such as `HTTP 200 OK` and the message which contains the actual object.

### 1.2.2 HTTP Metadata

HTTP requests and responses are coupled with metadata that describe these requests and responses. The HTTP metadata describes one of the following:

- HTTP Session - Describes the current HTTP session. For example, metadata "connection" describes if the web server should terminate the current HTTP session after this request / response or not.

- The Web Server - Describes the web server itself. For example, the metadata "Server" gives the name of the server.

- The Web Client - Describes the web client. For example. the metadata "User-Agent" is the user software that is requesting the resource

- Web Resource - Describes the web resource being served. For example, "Content-Length" is the length of the resource being served.

- HTTP Services: Caching, Content Negotiation - HTTP supports various services such caching at a web proxy or negotiation the form of the content. Some of the metadata fields describe the attributes of these services.

### 1.2.3 HTTP Methods

HTTP supports different request types. The following are the different types of methods that HTTP supports: `GET, HEAD, POST, PUT, DELETE, TRACE, OPTIONS, CONNECT`. The HTTP method that is responsible for fetching a web resource is `GET`. The HTTP method `HEAD` is used when the web client is only interested in fetching the metadata associated with the web resource and does not worry about the actual web resource.

## 1.3 Web Resources

The resources in the web can be categorized in the following three types.

- Static Resources: Static resources are the web resources do not change their form over a long time or based on the metadata presented in the request. An example of a static resource is a static image. A peculiar characteristic of a static resource is that the URL often points to a file name. For example, the URL `http://upload.wikimedia.org/wikimedia/google.jpg` points to a `JPEG` file.

- Dynamic Resources: A Web resource that is generated upon the request from a web client is called a Dynamic Resource. Personalized Facebook homepage is an example of a dynamic resource.

- Multiform Resources: Multiform resources are midway between static and dynamic resources. A multiform web resource exists in multiple different forms.

Based on the context presented by the requester, it changes its form. A webpage that changes its form based on the UserAgent is an example of multiform resource. The CNN homepage `http://www.cnn.com` is another such example. The CNN homepage changes its form as and when news arrive. But the URL that is used for retrieving the resource is still the same.

## 1.4 Conclusion

In this chapter, we studied the background the World Wide Web that is relevant to the contribution of the thesis. The following chapter talks about future of Internet research and a candidate future Internet architecture - eXpressive Internet Architectire (XIA) in which we implement our ideas.

# 2

# eXpressive Internet Architecture

The most common theme observed in the future Internet research is to move away from the host centric Internet. Computers communicating to one another was a cornerstone of the time when the Internet was born. Therefore, the IPv4 network on which the Internet is based addressed hosts in the network in order to establish communication between them. Thus, the Internet as we see today has evolved on the host based paradigm.

But, for the Internet users today it does not matter *who* serves the information. Users care about *what* information they receive. This shift in Internet usage has given rise to a new approach to evolve the Internet to a network infrastructure in which focal point is the *content* rather than *hosts*. This approach is generally referred to as Information Centric Networking. Figure 2.1 shows an illustration of Internet usage in an information-centric Internet vs today's host-centric Internet compared. An example of a proposed Information centric networking architecture is Named Data Networking(NDN).

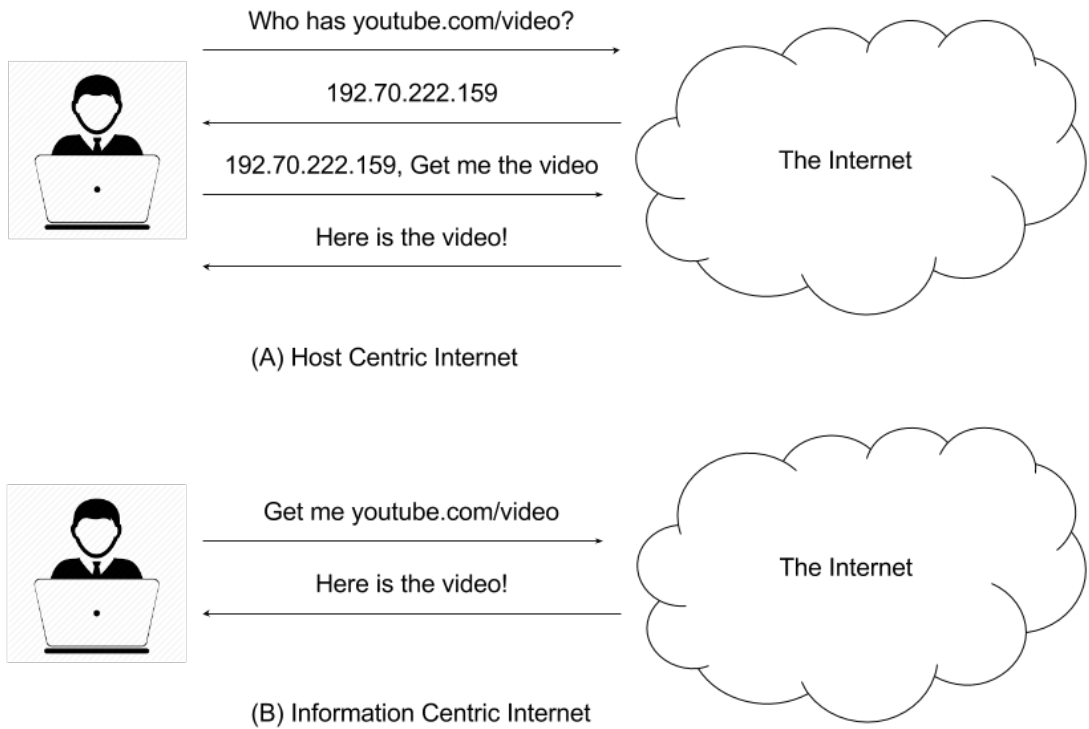

Figure 2.1: Host Centric Internet vs Information Centric Internet

## 2.1 eXpressive Internet Architecture (XIA)

eXpressive Internet Architecture is a candidate future Internet architecture that argues against elevating one particular communication *principal* type over others. A principal is a communication entity such as a host, a domain, a service or a specific content piece. If the network primarily supports communication with one particular principal type then communication with the other principal type inherently becomes difficult. We see in today's host-centric Internet that all the content requests first need to identify the host who is capable of serving the content request. Similarly, if we moved to an Information Centric Internet, it is not obvious how can we make hosts communicate with each other. XIA, thus, supports coexistence of multiple principal types.

Three key features of XIA are as follows:

- Multiple Principal Types

- Fallbacks

- Intrinsic Security

Let's go over these features one by one in order to understand XIA better.

### 2.1.1 Multiple Principal Types

As seen above, XIA supports coexistence of multiple communication principal types. The communication principals that XIA supports are hosts, services, administrative domains and content. These principals are identified by unique eXpressive identifiers called as XIDs. Corresponding to the four principal types mentioned above, their XIDs are referred to as HIDs, SIDs, ADs and CIDs.

### 2.1.2 Intrinsic Security

XIA's intrinsic security requirement mandates that a communicating principal must be able to prove itself. In other words, it should be possible for an entity that is communicating with a principal to verify the authenticity and integrity of the principal. XIA, thus, chooses the XID as such that they guarantee the authenticity and integrity of the communicating principal. For example, the host identifier or the HID is the secure hash of the host's public key. The content identifier or the CID is the secure hash of the content itself.

### 2.1.3 Fallbacks

In order to support evolvibility, it becomes important to address the issue of how network entities that do *not* understand a communication principal type deal with the principal. An analogous example in today's Internet could be as follows. If we move to IPv6 Internet, what if an intermediate network does not understand IPv6 and only understands IPv4? The intermediate nodes should not drop the IPv6 packets. In other words, the network still needs a way to forward packets even if it does not understand a communication principal. XIA supports this ability via the notion of fallbacks. Fallbacks allow XIA to specify multiple paths to the same principal. The entities that do not understand a particular path can take a different path for communicating with the principal type. All the different paths to the communication principal are combined into a network layer address that takes the format of a directed acyclic graph (DAG) as shown in Figure 2.2. The address is interpreted as follows - forward / route primarily based on CID, but if you don't understand CID, route based on IP address.

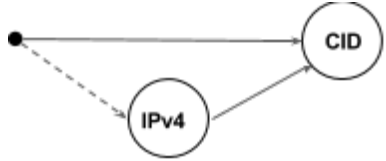

Figure 2.2: Fallbacks

## 2.2 Content Principal in XIA

With the content principal users can express interest in content irrespective of its location. Sending a content request fetches content from anywhere in the network. The request could go all the way back to the original publisher of the content or can be served from an in-network cache that holds a copy of the content. The API for content principal are as shown in Table 2.1. Content principal provides intrinsic security guarantees by choosing the cryptographic hash of the content as its identifier.

| Function | Description |
| --- | --- |
| getContent(socket, addr, buffer) | Retrieves the content specified by addr from network; addr contains CID and possibly a fallback |
| putContent(socket, content) | Registers the content as available. After making this call, the network knows how to fetch content. |

Table 2.1: API for Content Principal

Thus, all the network devices can verify authenticity of the content objects.

## 2.3 Conclusion

In this chapter, we studied the eXpressive Internet Architecture. We looked at the importation features offered by XIA - coexistence of multiple principal types, intrinsic security and fallbacks. Then we looked at how network entities can use the content principal to publish and fetch content irrespective of its physical location. This was the last introductory chapter in the thesis. The following chapters build on to ideas discussed in these two chapters and are the main contributions of the thesis.

# 3

# Content Retrieval Infrastructure

## 3.1 Motivation

Most of the applications today that need to deliver content reliably use TCP as the transport layer protocol. The motivating example in our case is the world wide web. TCP relies on communication from end-hosts to provide reliable content delivery. Relying on end hosts rather than the network for the reliable delivery information gives TCP the advantage that no explicit network layer feedback is needed to reliably transport content. This goes in accordance with the philosophy of a dumb network and smart ends and the end-to-end principle.

However, the rise of information centric networking shifts the focus from hosts to content. That poses us with the challenge of redesigning reliable transport for content centric architectures. The solutions that have been proposed, rely on client applications to fetch individual packets belonging to a object reliably. This approach of one-way reliability loses on important congestion control information that the content provider (a router cache or the original publisher) could have received from the client. An example of such information could be the congestion window size.

We, therefore, take a different approach to solve the problem of reliable trans-port of content chunks. We use coexisting content and service principals in XIA to allow us to apply TCP like congestion control and reliability approach to content centric networks. The content principal helps in locating the content and the service principal helps in delivering the content. We show that this approach of delivering content-over-service does not change the semantics of use and still allows both the publisher and the client to participate in a reliable content transport session resulting in a more efficient reliable content transport for content centric architectures.

## 3.2 Design Goals

As we have seen in the last chapters, XIA does not disregard the fact that the first class principal the world is moving towards is content. At the same time it allows for presence of multiple such principals simultaneously. In the new content retrieval infrastructure we aim to leverage the coexistence of service and content principals. We aim to solve problem the problem of reliable transport of content for content centric networks using the coexisting service and content principals.

While solving other problems, we aim to ensure that the benefits expected out of an information-centric architecture are still preserved. So, ICN features such as on path caching, content based routing should remain as they are.

Also, we envision that the cache on routers could extend in multiple dimensions. For example, cache could choose to store content at a remote location rather than in its local storage. Content eviction policies might change over time. Thus, it is important that the system we design is extensible in all the ways possible.

To summarize, the goals of new content principal handling are as follows:

- Support Opportunistic Caching.

- Support Reliable Transport for Content Objects.

- Support Extensible.

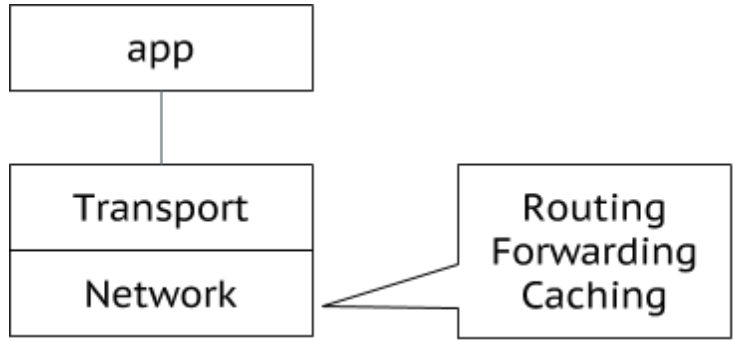

Figure 3.1: Old Network Stack for Content Delivery Architecture

## 3.3 Design Overview

As we have seen in chapter 2, cacheable content is addressed using CIDs. CID principal puts no limit on the size of the content chunk. Although desirable, this requirement implies that some applications will need the ability to transmit and fetch content reliably. How does it affect our design? Lets look at where various functionalities are implemented in XIAs network stack. The old network stack for content delivery architecture is outlined in Figure 3.1. Reliable transport service (streaming sockets) is implemented at transport layer in this stack. In order to transmit content reliably, we need a way to use reliable transport service. Therefore in our new design, we implement content delivery infrastructure primarily in an application while keeping content based routing as it is. The new network stack and the functionality split is as shown in Figure 3.2.

The fact that caching is moved to an application gives us following advantages:

- Extensibility

- Easy access to reliable transport API

We define a new application level entity called Xcache Daemon(xcached) that takes the responsibility of caching and serving content chunks. Xcached receives requests from client applications and translates them into socket calls as shown in Table 3.2. Xcached also takes care of multiplexing and demultiplexing of content requests and responses to and from various connected content applications.

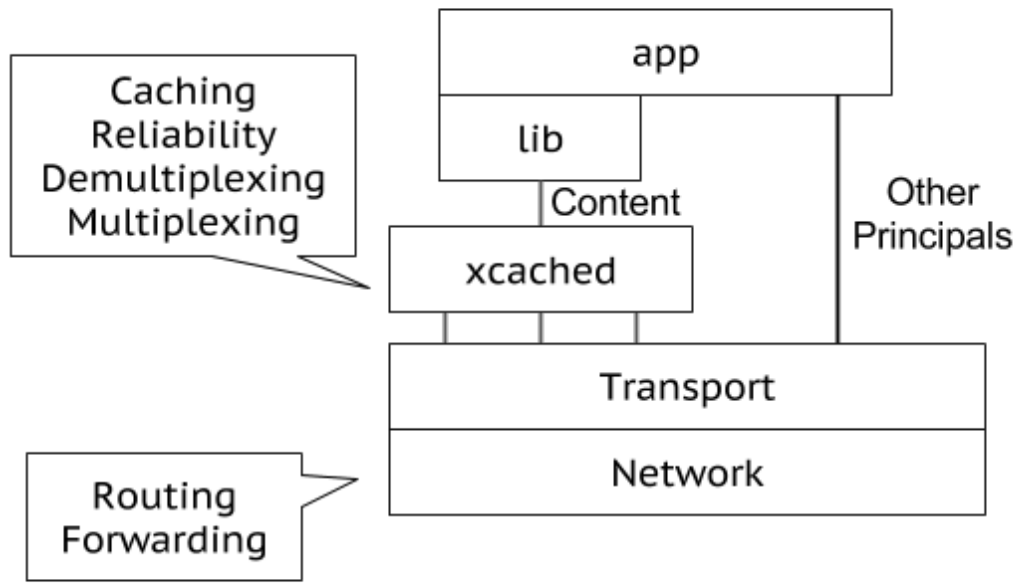

Figure 3.2: New Network Stack for Content Delivery Architecture

## 3.4 Xcached Architecture

Xcached is the center of all CID (as well as nCID as we will see in the later chapters) operations. So, it is important that it does not become the bottleneck in the content delivery infrastructure. It essentially means that xcached should not starve applications, it should process requests fairly and it should be performant. In this section we will look closely at the xcached daemon - how it has been architected and what are some of its extensibility characteristics.

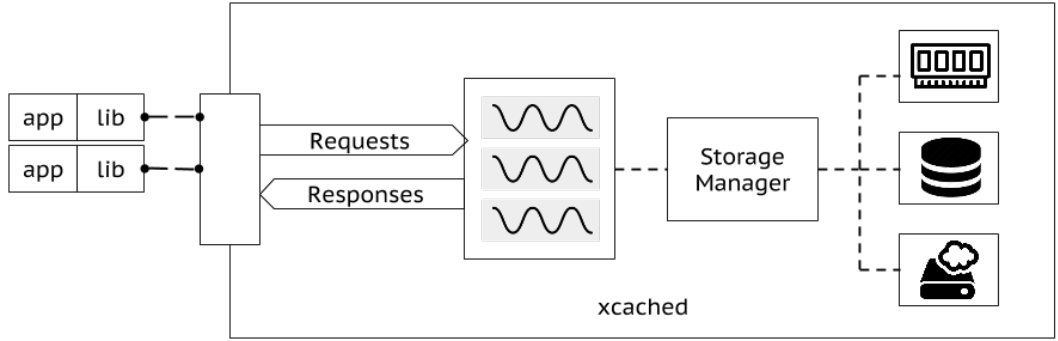

Figure 3.3: Xcached Architecture

Xcached is a mutli-threaded program that follows worker thread pool model in which the application facing controller puts the arrived requests in a queue and worker threads dequeue and do the work as and when convenient. Figure 3.3 shows the various modules in the xcache daemon.

### 3.4.1 XcacheLib

In order to hide communication details from the application, we have built a library that exposes the required APIs to the content applications. These APIs are described in greater details in section 3.7.

### 3.4.2 The controller

The controller is the application facing front end which takes requests from the XcacheLib. The communication between the controller and XcacheLib is over UNIX domain sockets. We use google-protobuf to parse and unparse the requests to and from the controller. It is possible that applications request for content that is cached locally. The controller processes such requests in the fast-path. I.e. It sends back the response to the application quickly. All the other requests which cannot be processed in the fast path are put in the requests queue. These requests are processed at some time in future by one of threads in the thread pool.

| Job | Details | No. of threads |
|---|---|---|
| Content Publish / Fetch | Storing content published by applications or fetching content from a remote source | Arbitrary |
| Content Eviction on Timeout | Content objects that get cached have a associated time-to-live period. This is the time period after which content object should become unavailable. | One |
| Opportunistic Caching | Opportunistic caching of content chunks by in-network devices | One |

Table 3.1: Jobs performed by threads in xcached

### 3.4.3 Thread Pool

Xcached thread pool consists of a set of worker threads, number of which can be configured at xcached startup. The thread pool consists of threads that perform one of the tasks shown in Table 3.1.

### 3.4.4 Content Stores

Different content systems have different characteristics: RAM allows fast retrieval of data but has limited size. Disk is slower than RAM store but can hold a lot more content. Thus storing content in RAM might be more desirable for use cases that need only smaller chunks whereas use cases that need to store big content chunks might prefer to store content on Disk rather than in RAM. In order to support these varying use cases, Xcached supports multiple different content storage methods. By default, it supports storing content in RAM, on disk and on a network attached storage device. Also, it is possible to compile new storage methods with xcached. Implementing a new storage method is as simple as extending following class and implementing the member methods.

```
class xcache_content_store {
        virtual store();
        virtual get();
}
```

3.4.5   Storage Manager

Since xcached has different storage methods, it is possible to organize content objects across these stores in different ways to serve different use-cases. Possible content placement policies could be round-robin, popularity based or size based. We implement a simple content placement policy which places content in RAM store until its full and then moves subsequent content to the disk store.

3.4.6   Content Eviction

If possible content stores run out of space, content must be evicted to make space for new incoming content. Content eviction policies govern which content chunks should be evicted on such an event from a particular content store. The content eviction policy that Xcache supports is LRU (Least recently used). Just like content stores, new content eviction policies can be compiled with xcache and associated with the content stores. Implementing new content store specific content eviction policies is just as simple as extending and implementing following class and associating it with a content store.

```
class xcache_eviction_policy {
        virtual store();
        virtual get();
        virtual remove();
        virtual evict();
}
```

## 3.5 Design Details

Xcached has three types of interfaces with the transport layer. In order to understand these interfaces better let's take a closer look at three network components that interact with XcacheD.

### 3.5.1 Content Server

Xcached acts as a content server on the publisher's end as well as when the in-network cache delivers content to the client. On receipt of content requests, the daemon needs to know what content is being requested. We define a new type of transport socket called "content server" socket which allows daemon to bind to all the content connections. This socket allows xcached to know for what content the request was received. The content server socket needs to do two unusual tasks which are significantly different from normal server sockets.

*Bind(Content \*)*

Since content server needs to listen to all incoming content request, it needs to bind to all content addresses that the provider has with it. Thus we need a notion of `Bind(Content *)`.

*AcceptAs(MyAddress)*

As the server side socket has been bound to many addresses, when an incoming request is received, the xcached needs to know for which address the request was received. In other words, the server needs a way to know what address is the source address for packets going out of it. Thus we define a new `AcceptAs` call that tells xcached for which content the request was received.

| API | What xcache does |
|---|---|
| XputChunk | StoreContent(), XaddRoute() |
| XgetChunk | Xconnect(CID Dag), Xrecv() |

Table 3.2: API Expansion

### 3.5.2 Content Client

The main responsibility of client side xcached is to establish a reliable transport session with a content provider and fetch content reliably. The content provider can be an in-network cache or it can be the end publisher. So, xcached's client side socket "connects" with a content provider and fetches the content. In other words, xcached's client socket "connects" with content rather than a "service".

### 3.5.3 Opportunistic Caching

The last context in which xcached comes into picture is opportunistic caching. When content providers serve content to clients over a reliable transport session, in-network devices need to intercept and cache content packets as they are traveling through them. The third interface that xcached has with the network allows xcached to sniff content packets flowing through it. We call this socket a content raw socket.

## 3.6 End to End Example

We have seen at high level how we moved caching and content serving functionality to xcached application. In this section we will walk through a detailed end to end example and see how clients establish a reliable transport session with a publisher and how intermediate caches cache the packets flowing through them.

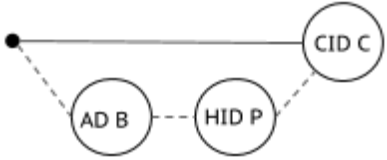

Figure 3.4: CID Dag Publisher

### 3.6.1 Step 1: `XputChunk(Content)`

The whole story starts with the publisher putting the content objects in the local cache (xcached) and publishing routes to these content objects. This lets the network know that unless cached at a better location, all the incoming requests for this content should be forwarded to this host. The end publisher publishes the chunk with the DAG address as shown in Figure 3.4.

### 3.6.2 Step 2: `XgetChunk(CID_Dag)`

- Client application calls `XgetChunk` with the desired DAG address as the argument. This call lets the xcached know that a client application is interested in fetching content pointed to by the DAG.

- Xcached tries to establish the reliable transport session with one of the many network entities who can serve the content by calling `Xconnect(CID-Dag)`. This `Xconnect(CID-Dag)` call serves two purposes: It acts as a content request as well the first packet of three-way handshake of the reliable transport session (SYN).

- Any network device that has the content chunk cached, accepts the GET / SYN packet and in effect tells xcached that there is a content request and it is the first packet of the reliable transport session.

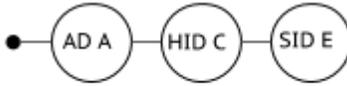

Figure 3.5: SYN Source Address

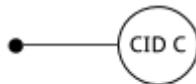

Figure 3.6: Published Content Chunk Address

- Call to `XacceptAs` made by xcached then returns with the address of content that was requested by the client. Calling `XacceptAs` results in generation of SYN-ACK. The return of `XacceptAs` also means that the three-way handshake was completed by the xcached that provides the content.

### 3.6.3 Addresses

The SYN packet that xcached on the content provider received, has the source address that looks like address in Figure 3.5.

The SID in the source address represents the ephemeral reliable transport endpoint that xcached on client had created. This address acts as a way back to the client. Content server socket completes the three way handshake by accepting the connection with source address as in Figure 3.6. Once the connection has been accepted, content provider serves the content over the established reliable transport session.

### 3.6.4 Opportunistic Caching

The challenge in opportunistically caching the content objects is that the xcached on the in-network cache is not an active participant in the reliable transport session. We solve the problem of opportunistic caching by forwarding all the content packets to xcached and then reassembling the chunk by peeking into transport header. Following are the steps that take place on the intermediate cache when content object is to be cached.

- While content is being served from the content provider to content client, it gets caught by the CID raw sockets on Xcached's running on intermediate network devices. The CID raw socket forwards all the packets which have the primary intent as CID to the xcached running.

- Looking at the first packet in the transport session (SYN/ACK), the forwarding engine needs to know if the content object should be cached or not. The logic for this policy is implemented in xcached. Based on the policy, xcached takes a decision and either sends "Yes, cache" or "No, don't cache" decision to the forwarding engine

- Packets belonging to all content chunks for which the policy decision is "Yes, cache" are forwarded to xcached.

- Xcached peeks into the transport header and reassembles a content object. Once reassembled, now that particular xcached also acts as the content provider for the chunk.

## 3.7 API Details

With the movement of caching and content handling to xcache application, we also redefined the application interfaces for publishing and fetching content over XIA. The application is not responsible for establishing the reliable transport session. Xcached does it on behalf of the application. This simplifies the application design and also gives the xcached the ability to cache content that was requested by the application. In this section, we will go over the APIs that XcacheLib exposes to the content applications and see an example of a sample content publisher as well as a client application.

- `int XcacheHandleInit(XcacheHandle *h)` - Creates a connection with `xcached` and fills in the opaque structure `XcacheHandle`. `XcacheHandle` acts as a context for the rest of the APIs.

- `int XcacheHandleDestroy(XcacheHandle *h)` - Destroys the the handle created by the call `XcacheHandleInit`. Applications that call `XcacheHandleInit` must call `XcacheHandleDestroy`.

- `int XfetchChunk(XcacheHandle *h, void *buf, size_t buflen, int flags, sockaddr_x *addr, socklen_t addrlen)` - This function lets applications fetch content chunk that has the DAG address `addr` of length `addrlen`. The Xcache context `h` must be initialized by calling function `XcacheHandleInit`.

- `int XputChunk(XcacheHandle *h, const void *data, size_t length, sockaddr_x *addr)` - `XputChunk` allows applications to publish chunks to the network. A call to `XputChunk` publishes a chunk which has data pointed to by `data` and of length `length`. The address of the published chunk is returned in the address `addr`.

- `int XregisterNotif(int event,`
  `void (*func)(XcacheHandle *, int event, sockaddr_x *addr, socklen_t`
  `addrlen))` - Xcache allows applications to listen for certain "notifications". Examples of these notifications include chunk eviction notification, chunk arrival notification. A call to `XregisterNotif` registers a handler for a particular notification. Applications can either spawn a separate thread for handling notifications by calling `XlaunchNotifThread` or they can look for received notifications by checking received data on socket returned by `XgetNotifSocket` and calling `XprocessNotif` if appropriate.

- `int XlaunchNotifThread(XcacheHandle *h)` - This function launches a notification listener thread. If `xcached` sends a notification on notification socket, the function calls registered handlers if appropriate.

- `int XgetNotifSocket(XcacheHandle *h)` - This function returns the socket on which `xcached` sends back the notifications.

- `int XprocessNotif(XcacheHandle *h)` - If the `xcached` notifcation socket has any incoming data, applications can call `XprocessNotif` to invoke the registered handlers.

## 3.8 Sample Applications

In this section, we will see how applications can use above mentioned APIs in their code. Section 3.8.1 shows a self-explanatory example of a content server application. Whereas, section 3.8.2 shows a self-explanatory example of a content client application.

### 3.8.1 Content Server Application

```
int main(void) {
  XcacheHandle xcache;
  sockaddr_x info;
  ...
  XcacheHandleInit(&xcache);
  ...
```

```
    XputChunk(&xcache, data, datalen, 512, &info);
    ...
    XdetroyChunk(&xcache);
}
```

### 3.8.2 Content Client Application

```
int main(void) {
    XcacheHandle xcache;
    sockaddr_x info;
    ...
    XcacheHandleInit(&xcache);
    ...
    XfetchChunk(&xcache, buf, 1024, XCF_BLOCK, &info, sizeof(sockaddr_x));
    ...
    XdetroyChunk(&xcache);
}
```

## 3.9 Conclusion

In this chapter, we saw how we moved content caching and serving to an application daemon `xcached`. In contrast to the old design, we used reliable transport to deliver content. We then looked at the new API and how we can use it to write content applications in XIA.

# 4
# Fetching Named Content in XIA

## 4.1 Motivation

In order to effectively utilize in-network caches, it is important that the resources in the web are cacheable. Cacheability of the resources really depends on its reusability. Static resources (objects images, video chunks) are probably the best candidates for utilizing in-network storage. But, the majority of web content falls in the other two categories.

It can be argued that dynamic resources can benefit the least from of in-network caches. The primary reason being that dynamic resources are generated upon the request arrival and hence are highly non-reusable. This, clients can thus

But the resources that are of interest to us are the multiform resources that we saw in chapter 1. Recall that these resources exist in multiple forms at a same time or at different times. Clients *choose* a representation that suites their need the best. They are reusable for the very reason that upon sending the same request multiple times, they respond with the same content. What can we do to make these multiform resources cacheable at the same time preserve their multiformity?

### 4.1.1 Human Readable Names for Multiform Resources

We have seen that web relies on human readable names for identifying multiform resources and in general all the web resources. Even though cryptographic identifiers have inherent security properties, what makes human readable names a preferred choice?

The first obvious reason is that the names are more understandable. Users establish trust in human readable Identifiers such as "facebook.com" much easily than in random cryptic hex string such as "0ABF1866BD7182...".

The second reason is that multiform resources change their representation. For example, content associated with the resource identified by "http://www.cnn.com" changes often. So, as time passes multiform web resources change their representation. Also, time is not the only dimension along which these resources change their representation. Another such dimension could be UserAgent. With the advent of smart-devices, number of platforms from which web resources are accessed has been increasing like never before. That poses us with the challenge of presenting the same resource in different forms based on the platform that the user is requesting the resource from.

CIDs have the disadvantage that an identifier can only point to a particular representation. It also implies that it is not possible for us to allocate an identifier for content that we dont know in advance. So, crypto IDs do not support the property of late binding.

## 4.2 Locating Content using Human Readable Names

In the last section we saw the reasons why the world wide web identifies and locates resources by human readable names. Lets see the possible options to locate content using human readable names in eXpressive Internet Architecture.

- Keep a DNS like mapping system that maps names to CIDs

- Locate content directly by human readable names

### 4.2.1 Problems with DNS like mapping system

In this approach, we would need a naming system that maps human readable names to CIDs. Users first query the naming system to get the CID and then request the CID. The naming systems could well be established at organizational levels like today. Although scalable, this approach suffers from following issues.

Size of the mapping system: Firstly, the size of the mappings that the system needs to maintain is directly proportional to number of cacheable objects that the publisher has as opposed to number of hosts in todays Internet. It is evident that number of objects outnumbers number of hosts by orders of magnitude. So, we would need really huge mapping system.

Number of roundtrips: In order to fetch a CID for a particular name, the web user would need to make several roundtrips to and from the naming system. Depending upon the model used by the naming system, the consumer could take from one to several roundtrips before it can know the CID corresponding to the content name.

Security Issues: Once the consumer receives CID corresponding to a human readable name why would the consumer believe that mapping between the human readable name and the CID is authentic? Hence, the content retrieval architecture must address these security issues somewhere. However, that loses the whole point of having cryptographic identifiers. Remember that one important property of CIDs is that they are self certifying.

Because of the issues mentioned above, it becomes unwise to use a naming system for locating content by name. Therefore, we take the second approach of directly addressing content by human readable identifiers. The following sections describe the new principal type that we define and address the security concerns raised by human readable names.

| API | Details |
| --- | --- |
| XputNamedContent(Buffer, Name, Certificate, Signature) | Publishes a named content chunk with the corresponding signature and the certificate |
| XgetNamedChunk(Name, Certificate) | Fetches a name content chunk and verifies the authenticity and integrity using the public key certificate |

Table 4.1: nCID Principal Semantics

## 4.3 nCID Principal Type

### 4.3.1 Definition

In order to effectively eliminate a huge naming system and support the multiform content on the world wide web, we define a new principal type which allows locating content directly by human readable names. Just like CIDs, the content requested by the consumer using nCID is retrieved from anywhere in the network. The XID type nCID is defined as follows:

`nCID = hash(Human Readable Name + Publisher's Public Key Fingerprint)`

nCID is thus content with a human readable name that is verified by a publisher.

### 4.3.2 Semantics

The nCID principal allows users to retrieve content identified by human readable names from anywhere in the network. Table 4.1 shows the APIs that nCID supports. Similar to CIDs, sending content request for nCID type using `getNamedContent()` initiates a transport session with any in-network cache or the original publisher. This reliable transport session is then used to deliver content to end consumers.

The `publishNamedContent(name, signature, public key fingerprint)` call tells the network that the content identified by a human readable name is available with the publisher. It is verified by a private key corresponding to the public key passed as an argument. Xcache expects that the signature passed an argument is generated in a certain way (we elaborate upon it in the following section). The public key fingerprint and the signature is used by in-network devices and the endpoints to perform security checks.

## 4.4 Security Issues with Named Content

### 4.4.1 nCID Security Requirements

XIAs intrinsic security requirement makes it mandatory for the XIA prinicpals to provide integrity and authenticity for the communication operation. CIDs provide the integrity and authentication by defining identifiers as the hash of the content so that when the consumer receives the content and the CID, it has a reason to believe that the content has not be tampered with and has been received from an authentic publisher.

We argue that satisfying following four requirements provides exactly these guarantees for nCID principal type.

1. Ability to Verify Binding between Name and Content

   The consumer and in-network devices should have a reason to believe that the content and name are tied together. CIDs provide this guarantee inherently. For nCIDs, we rely on public key infrastructure to provide the guarantees.

2. Ability to Verify Content Integrity

   The consumer should have a reason to believe that the content has not been tampered with.

3. Ability for Caches to Verify All These at Minimum Cost

   The in-network caches need to perform these security checks before they can cache a copy of the object. Applications can use different trust models. It would be inefficient for in-network caches to verify authenticity, integrity properties of the content chunk based on the application specific trust model. So, we need an application agnostic security model to work with.

4. Protection against Content Poisoning

    Since there is no absolute binding between the name and the content, an attacker can claim that certain content can belong to a particular name. Our security model must prevent such an action.

### 4.4.2 Security Model

Previous work argues that in order to provide security guarantees it is sufficient to bind any two pairs between name, content and publisher. We choose to bind name-content and name-publisher pairs.

In order to understand how our security model functions let us look at some important chunk headers that nCID chunk contains.

Name-content binding is provided by the digital signature that the publisher generates at publish time. Name-publisher binding is provided by the nCID itself. Following equations show how these bindings are created by the publisher.

```
nCID = hash(Content Name, Publisher's Public Key Fingerprint)
Signature = Encrpyt_with_Publisher's_Private_Key (Content Name,
                                Content Data)
```

As long as the consumer knows about the name of the content and the original publisher's public key fingerprint, it can always generate a content request. So, we expect that some high level entity (such as a TLS connection) delivers these two parameters to the end consumer. In the next chapter about URLs, we see how sophisticated URLs for nCID can be used to serve this information.

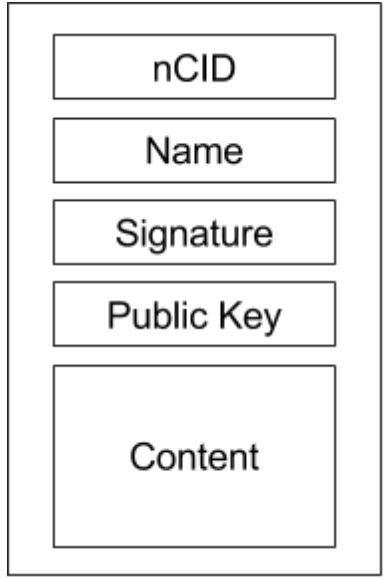

Figure 4.1: nCID Content Chunk Structure

nCID intrinsic security checking is a two-step process in contrast to one-step process in case of CIDs. The entity that needs to verify authenticity and integrity of the content chunk must first fetch the public key that was used to verify the content. The content chunk contains a pointer to the public key chunk.

Once the public key has been received, the consumer checks if nCID matches the hash of name and publishers public key fingerprint. It then decrypts the signature with the same public key. The decrypted signature is matched against name-content pair. If both these checks succeed, then consumer can safely believe that content is authentic and has not been tampered with.

No matter what trust model the application uses, all the in-network devices need to perform only one public key fetch and the two checks for nCID and signature. Thus, the time required for verifying security properties is constant. Besides, it requires at most only one content chunk fetch. This satisfies our third requirement of verifying security properties at minimum cost.

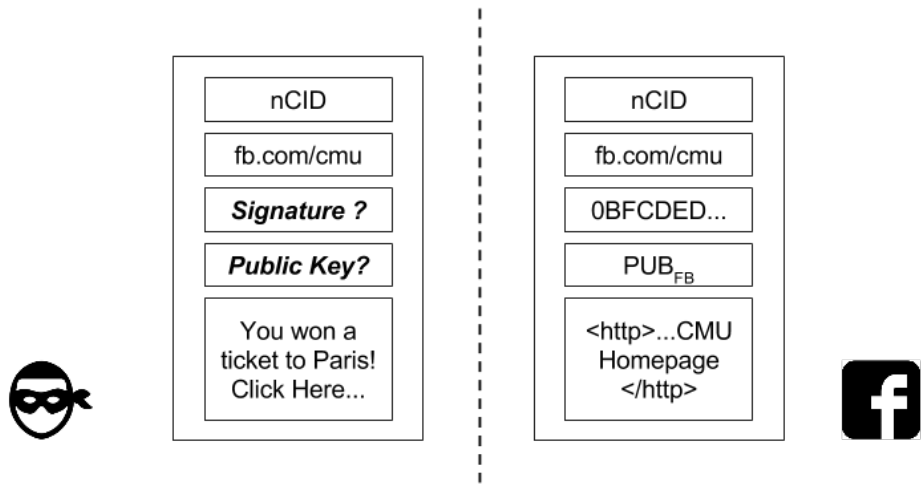

Figure 4.2: Content Poisoning

### 4.4.3 Protection against Content Poisoning

Content poisoning is an attack in which the attacker claims that a certain malicious content belongs to certain name that has already been published in the network. Figure 4.2 shows the motivating example.

In this example, the original publisher facebook.com has published a nCID chunk named "fb.com/cmu". Facebook has verified the chunk and put its signature in the chunk header. Now, an attacker wants to claim that certain spoofed content actually belongs to the name fb.com/cmu. What possible options he has to fill in signature and the public key?

- **Option 1: Public Key = Facebooks Public Key**

  If attacker uses facebooks public key itself then he cannot generate the corresponding signature. So, this option is not really possible.

- **Option 2: Use my own key**

  Lets say the attacker uses his own key to generate the signature. In that case,

even though attacker successfully plants a spoofed signature in the chunk he breaks the nCID check.

So we have seen that the attacker has no way of successfully filling in the signature and public key field pairs. Content poisoning is thus not possible in our security model.

## 4.5 Conclusion

We defined a new content principal for XIA that allows us to directly address the content by human readable names. We argued that such a principal best suites the "multiform web resources". However, such a content addressing system faces the issue of content authenticity and content integrity. We defined security models that address these authenticity and integrity issues to provide us an alternate secure content principal.

# 5

# URLs for Content in eXpressive Internet Architecture

## 5.1  Universal Resource Locators (URLs)

The resources in the world wide web are identified by **Universal Resource Identifier** often acronymed as **URIs**. URIs allow identifying resources uniquely thus giving hosts the ability to express interest in specific resources. The most common form of URIs is **Universal Resource Locators or URLs**. URLs, in addition to uniquely identifying a resource also specifies a mechanism to retrieve the resource. The URLs in general look like the following:

$$\texttt{protocol://host/path}$$

The *protocol* field specifies the scheme that should be used to retrieve a representation of the resource. The most popular resource retrieval scheme in the web is 'HTTP'. Other possible schemes are 'FTP', 'file', 'data'. The *host* part is the network location of the resource. It can be an IPv4 address or a domain name that the DNS can resolve to an IPv4 address. A *host* can own multiple resources. The *path* is the specific location of the resource on the *host*.

Figure 5.1: An example CID DAG address

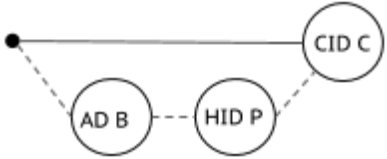

## 5.2 URL Design Goal

The goals of URL design for CIDs and nCIDs are two-fold.

Firstly, the web resources refer to other web resources all the time. For example, the HTML web pages contain URLs of other web resources. In the previous chapters we have seen that the CIDs can be used better to represent static web resources. On the other hand, nCIDs are a better fit for multiform resources. Since, in this thesis we use CIDs and nCIDs to represent web resources, we need a method of pointing to these web resources.

Secondly, the advantage that nCIDs have over the CIDs is that they are addressable by human readable names. Hence, constructing nCID URLs is a fairly understood problem. But the goal of URL design for CIDs is that the constructed URLs should be able to directly refer to the content avoiding name lookups.

## 5.3 URLs for CIDs

We propose following format for CID URLs.

$$\texttt{cid://serialized-cid-dag}$$

In order to understand the process of DAG serialization, let us take an example of a CID DAG address as shown in Figure 5.1.

We follow the following process to make a serialized version of the DAG:

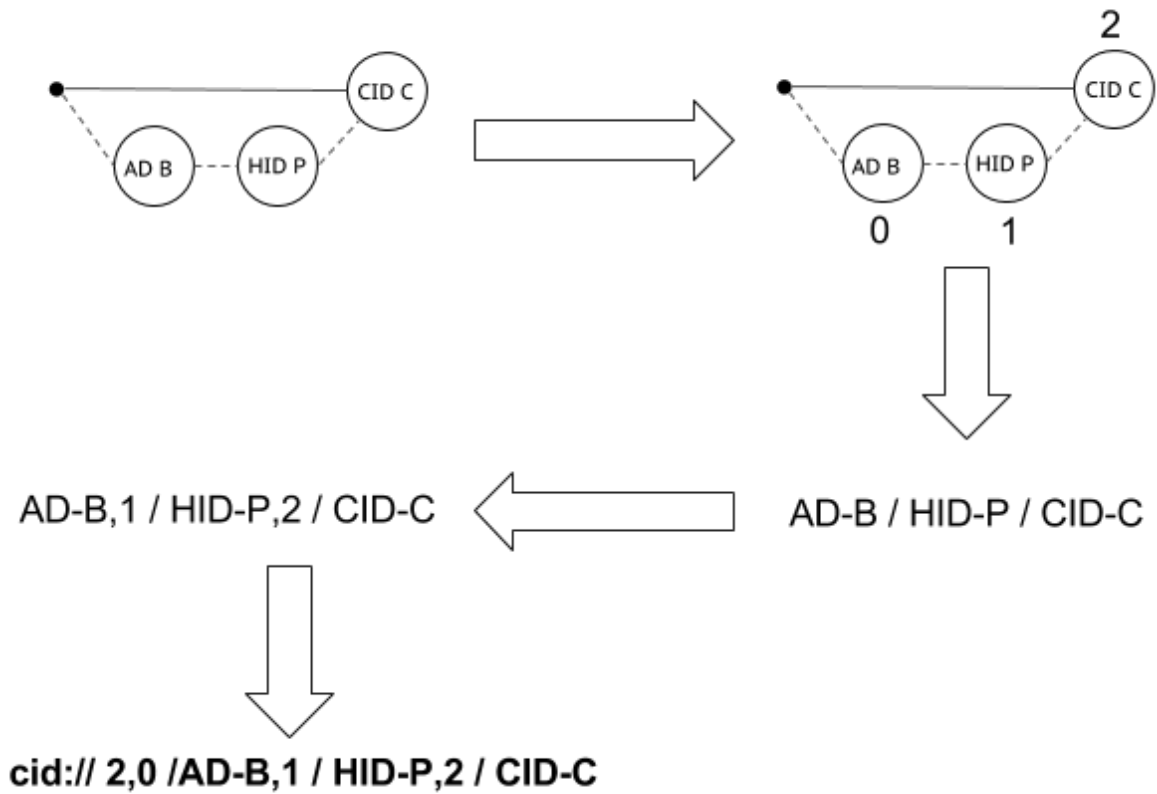

Figure 5.2: CID DAG Serialization

1. Number all the nodes starting with zero and excluding the source node.

2. List all the nodes in the order they are numbered from zero to maximum.

3. For each node, associate the destination node number for all the *outgoing* edges from that node.

4. Prepend the output of the last step with the outgoing edges from the source node.

This scheme results in an URL for CIDs that looks like this

**cid://2,0/AD-B,1/HID-P,2/CID-C**

The URLs that are created by this scheme are highly expressive. You can see that *any* DAG can be expressed by this method. That allows us to use protocols other than CIDs too. For example, we could create an URL for representing a service S as follows.

$$sid://2,0/\text{AD-B},1/\text{HID-P},2/SID\text{-S}$$

These URLs, since they directly address the content, avoid name lookup completely. With these URLs we can now represent static resources in the web.

## 5.4 URLs for nCIDs

We could readily use the URL scheme that we discussed in section 5.3. But, the disadvantage of such an approach is that the URLs created are highly unreadable. Also, as we see in more detail in the following section, nCIDs are formed using *attributes* that define the representation of the content chunks. The goal of URL design is to include those attributes in the URL. Since nCIDs are directly derived from human readable names, we define a URL scheme which results in URLs as we see today and yet avoids the need of name look-ups completely.

### 5.4.1 Addresses and Instance Addresses

We have seen that nCIDs are most useful for representing the multiform web resources. The peculiar characteristic of multiform resources is that they change their form based on certain *attributes*. In essence, all the representations of the same resource share the same *address* but are uniquely identified when the address is coupled with the *attributes*. We call the attributes that locate a certain representation of a resource the *instance address*.

As a motivating example, let's take a case of multiform resources identified by URL *http://wikipedia.org/google*. It is easy to see that when this web resource is requested from a desktop it takes a certain representation. While the same resource, if requested from a mobile device, takes a completely different form. So, even though the two objects are totally different in terms of their content, they do share an address. This identity is their address. To sum up, address of a multiform resource allows us to identify a resource and instance address helps us locate the derived representation of the resource.

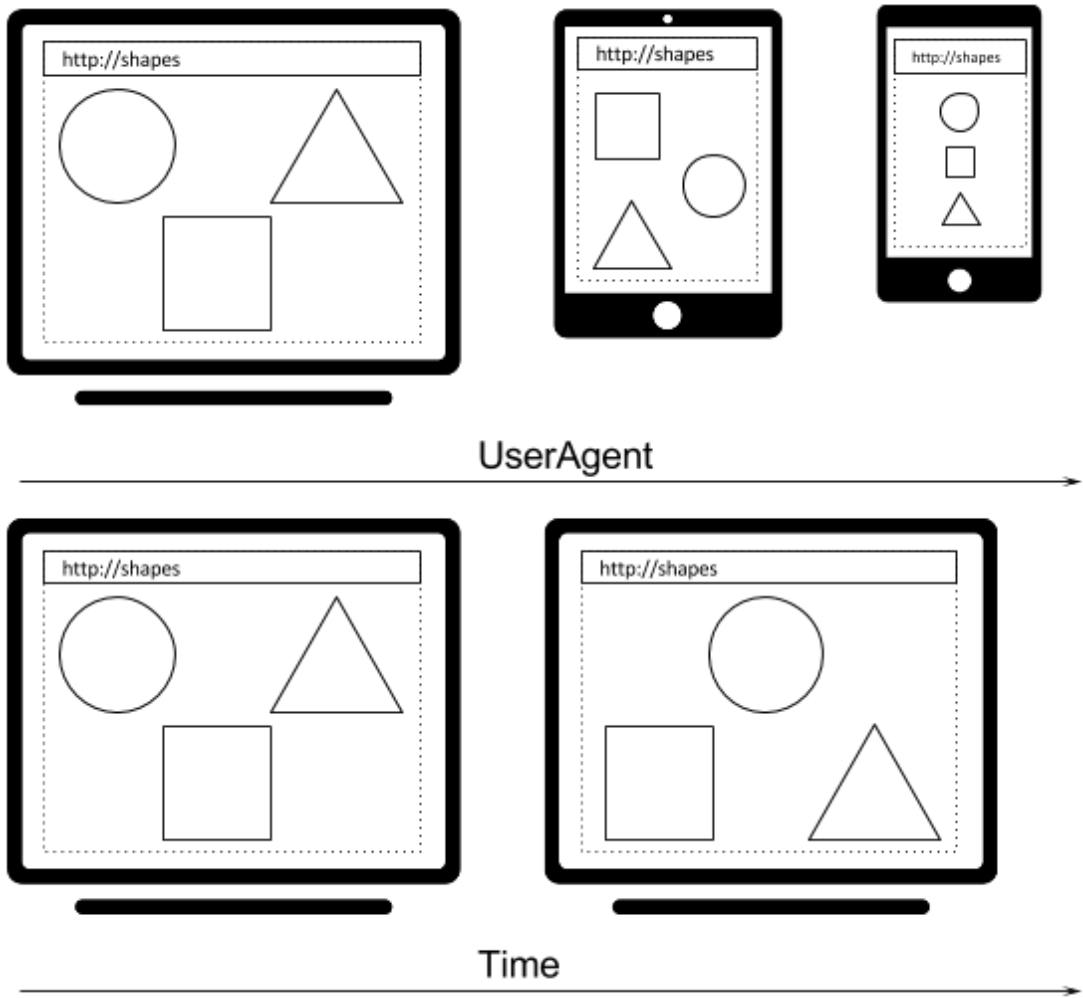

Figure 5.3: Multiform resource changing its form

| Name | Details | Mandatory |
| --- | --- | --- |
| PubCert | DAG Address of the Publisher's Public Key Certificate | Yes |
| Version | Version of the Content Chunk | No |
| UserAgent | HTTP UserAgent | No |

Table 5.1: List of Locators

### 5.4.2 nCID URL Design

Based on the learning from section 5.4.1, we propose the following format for the URLs of nCIDs.

$$\text{ncid://address/locator1=value1\&locator2=value2\&...}$$

With respect to the motivating example shown in Figure 5.3, the address of the content is *content.facebook.com* whereas an locator is *UserAgent=Android*. Table 5.1 shows the list of possible locators. Our security model enforces us to mandate either the implicit or the explicit existence of a locator called *certificate*. The locator *certificate* is essentially the pointer to the public key certificate that would be used to verify authenticity and integrity of the named content chunk.

## 5.5 Conclusion

In this chapter, we defined URL schemes for CID and nCID type content chunks. With CID URLs, we provided a way to point to CID chunks while avoiding name lookups. The nCID URL scheme allows us effectively express a link to a nCID resource. Separating addresses and locators gives web users the ability to choose specifically the representation that suites its needs the best.

# 6
# Conclusion

The old implementation of the content principal did not support reliable transport of content objects. The ability to reliably deliver content objects is crucial for the World Wide Web. We thus implemented content principal implementation to an application called XcacheD. Moving content principal implementation to an application was not a trivial task. It involved redefining interfaces with the network stack. We carefully studied possible approaches and implemented the content principal handling in Xcached application.

In order to fetch content objects directly by human readable names, we defined a new XIA principal type - nCID. We addressed the authenticity and integrity issues of the new principal type. The new principal type allowed the clients to avoid name lookups for fetching content chunks.

We classified the web resources into three categories: static resources, dynamic resources and multiform resources. XIA's different communication principals supported these different web resources. We argued that the static resources can be well represented with CIDs, the dynamic resources can be well represented with SIDs and the multiform resources can be well represented with nCIDs.

In order to effectively reference these various resources, we defined a URL scheme. We defined a generic URL scheme to map any XIA address to serialized character string. This scheme allowed us to represent addresses of static and dynamic resources. We then defined URL format for nCID principal that allowed us to address the multiform resources.

With the above contributions, we modeled the World Wide Web on eXpressive Internet Architecture.